# Post-Pandemic Hybrid Work in Software Companies: Findings from an Industrial Case Study

Ronnie de Souza Santos
University of Calgary & CESAR
Calgary, AB, Canada
ronnie.desouzasantos@ucalgary.com

Willian Grillo
CESAR
Recife, PE, Brazil
wng@cesar.org.br

Djafran Cabral
CESAR
Recife, PE, Brazil
dac@cesar.org.br

Catarina de Castro
CESAR
Recife, PE, Brazil
ccmk@cesar.org.br

Nicole Albuquerque
CESAR
Recife, PE, Brazil
nicole.pereira@cesar.org.br

Cesar França
CESAR
Recife, PE, Brazil
franssa@cesar.org.br

## ABSTRACT

*Context.* Software professionals learned from their experience during the pandemic that most of their work can be done remotely, and now software companies are expected to adopt hybrid work models to avoid the resignation of talented professionals who require more flexibility and work-life balance. However, hybrid work is a spectrum of flexible work arrangements, and currently, there are no well-established hybrid work configurations to be followed in the post-pandemic period. *Goal.* We investigated how software engineers are experiencing the post-pandemic hybrid work landscape, aiming to understand the factors that influence their choices between remote and in-office work. *Method.* We explored a large South American company by collecting quantitative and qualitative data from 545 software professionals who are currently navigating diverse hybrid work arrangements tailored to their individual and team requirements. *Findings.* Our study revealed an array of factors that significantly impact hybrid work within the software industry, including individual preferences, work-life balance, commute time, social interactions, productivity, and more. Team dynamics, project demands, client expectations, and organizational strategies also play an important role in shaping the complex landscape of hybrid work configurations in software engineering. *Conclusions.* In summary, the success of hybrid work models depends on balancing individual preferences, team dynamics, and organizational strategies. Our study demonstrated that, at present, there is no one-size-fits-all individuals approach to hybrid work in the software industry.

## 1 INTRODUCTION

The COVID-19 pandemic became one of the most devastating events in the history of humankind. As the virus spread across the globe, it forced society to find alternative ways to keep functioning [42]. In particular, remote work became one of the only alternatives for organizations. In this process, many professionals had to deal with working challenges when adjusting their homes to unconventional daily routines, e.g., working in improvised home offices, sharing the space with family members, or juggling work with childcare. [4, 6, 9, 29].

After months of restrictions, when the pandemic was under control, many companies started asking employees to return to their offices; however, professionals adapted to the challenges experienced at the beginning of the pandemic, and the idea of returning to an office routine was no longer attractive [5]. In the software industry, the plans to return to the office triggered different feelings among software engineers. Some were keen; others were trepidatious, but most were convinced by personal experience that their work could be done remotely. Therefore, they expect to continue working from home some or all the time [30, 35].

Software professionals have several reasons to maintain remote work structures, including more time for personal responsibilities (e.g., parenting), no commuting [18, 33], and more opportunities for people from underrepresented groups [12, 13]. Thus, many of them, when pressed to return to the office, have quit in favor of more flexible jobs [1]. Consequently, many software companies are adopting hybrid work to allow employees to work partly remotely and avoid the resignation of talented professionals[23].

The software industry has a history of geographically distributed teams and appears to be more accepting of new work arrangements [14, 33]. However, hybrid work differs from global software development, which is mainly focused on an onshore organization outsourcing to offshore partners [11]. Hybrid work is a spectrum of flexible work arrangements [37], in which employees usually belong to the same organization and mostly live in the same region where the office is (i.e., no outsourcing); they simply choose not to work in the office every day [11, 14].

In this study, we investigated how software professionals are experiencing post-pandemic hybrid work by exploring the case of a large South American software company that adopted a hybrid-work model, i.e., professionals can decide if and when they are going to the office based on personal, team, or project needs–something internally defined as *full-flexible hybrid work*. To address this topic, we focused on the following research question:

**Research question:** *How are software professionals dealing with post-pandemic hybrid-work?*

In line with this research inquiry, we explored the characteristics of professionals who regularly work from the office, the preferences of software professionals concerning the structure of hybrid work, and assessed the impact of post-pandemic hybrid work on their perceived productivity.

From this introduction, our study is organized as follows. In Section 2, we discuss existing studies about hybrid work in software development. In Section 3, we describe how we conducted the survey, while Section 4 presents our findings. In Section 5, we discuss the implications and limitations of our study. Finally, Section 6 summarizes the contributions of this study.

## 2 BACKGROUND

The concept of hybrid work is not new. Hybrid work has been addressed in the literature since long before the pandemic [8, 21, 36]. In the early 2000s, studies were reporting experiences in organizational settings where employees were working both at home and also in physical spaces (i.e., offices), using information and communication technology to organize workloads and communicate [21].

The definition of hybrid work, on the other hand, was never uniform, as the term hybrid is commonly used to refer to any environment that combines working from home and working in the office, that is, integrating face-to-face or using computer-mediated interaction [8, 36]. Even today, when hybrid work has been popularized across several industries after the pandemic, this concept is not uniform and has been used to characterize a spectrum of work arrangements that lies between working remotely and working in an office [22, 28, 37].

One approach to structuring post-pandemic hybrid work arrangements within the software industry involves distinct dimensions, including [37]: office mode, office-first, office-remote mix (flexible and location-independent work), remote-first, and remote mode. *office mode* means a more conventional work environment where the majority of team members are predominantly office-based. *office-first* prioritizes office work with occasional planned remote days. In *office-remote mix*, professionals benefit from the flexibility and the ability to work from various locations, resulting in varying office attendance. *remote-first* emphasizes remote work, with software professionals spending limited planned days in the office. Finally, *remote mode* indicates a fully dispersed team engaging primarily in remote work.

In software engineering, since software professionals are demanding more flexibility in the post-pandemic period [30], hybrid teams have become the focus of many studies [7, 10, 11, 25, 26, 35, 37, 41] aiming to understand the effects of this practice in the software development. Before the pandemic, there were two main challenges associated with hybrid work. First, the management of working practices, in particular, coordination and integration of tasks [21, 36]. Second, the decreased levels of interaction among co-workers in hybrid environments [8, 21]. Nowadays, the list of challenges has increased considerably, as many employees are not happy about losing the benefits they acquired when working fully remotely, which now might be affected by regular visits to the office [3, 24].

Post-pandemic hybrid work produces several benefits and limitations to software professionals [7, 11–13, 26, 32, 35]. On the one hand, hybrid work allows software professionals to maintain high levels of flexibility, more autonomy, autonomy and an increased number of job opportunities experienced in fully remote work arrangements [32, 35] while also providing software teams with more diversity [12, 13]. On the other hand, software teams are facing problems associated with communication, coordination, and cohesion [11, 15, 35], processes that are essential for any high-performance team [16]. In this sense, software companies are expected to explore and understand the impacts of hybrid work on their teams and identify how the different configurations of this new work arrangement can be effectively applied in their workplace [7].

## 3 METHOD

Case studies are commonly applied in software engineering to investigate a contemporary software engineering phenomenon by exploring multiple sources of evidence in a real-life setting, e.g., a software company or a small number of software teams [34]. Case studies provide a systematic approach to exploring events, collecting and analyzing real-world data, and reporting findings that can be used to increase the understanding of situations and inform industrial practice [2].

In this research, we conducted a case study because this method is consistent with the phenomenon under investigation, as we are interested in understanding how software professionals are experiencing hybrid work and what their perspectives about going back to work in the office, e.g., voluntarily or following project or employee demands. This is a relevant contemporary discussion in the software industry because there is currently no consensus about how software companies should implement hybrid work, and previous studies have demonstrated that this work arrangement generates conflicting outcomes for professionals and businesses.

### 3.1 Selecting the Case

The company selected as our case study has been in business since 1996, developing software solutions in several domains, including finance, telecommunication, government, manufacturing, services, and utilities. More than 1,200 professionals work in this company, and more than 70% of them work directly in software development activities in 50 different software teams.

These teams are composed of software professionals from various backgrounds, both technical (e.g., programmers, QAs, designers) and personal (e.g., diversity of gender, diversity of ethnicity). These professionals are familiar with most software development methods (e.g., Scrum, Kanban, and Waterfall), processes, tools, and techniques for software development as they develop software for global clients, including North America, Latin America, Europe, and Asia.

This organization was intentionally selected because it provides an ideal context for exploring post-pandemic hybrid work models as during the COVID-19 pandemic, the company required that employees work from home indefinitely, and now, after pandemic restrictions were lifted, professionals were given the option to work in a flexible hybrid arrangement, with their presence in the office depending on their preferences and project requirements. However, the company is observing varying levels of acceptance and resistance among software professionals with post-pandemic hybrid work arrangements. Consequently, the organization is actively working to define one or several optimal hybrid work setups that align with the diverse needs of employees, clients, and the overall business strategy.

According to the company's executives, when the research took place in April 2023, they needed to establish a hybrid environment that works and effectively supports their business needs. In their setting, since they have a business structure based on developing solutions for external clients, they need these clients to experience the organizational culture when they visit the office, and it is difficult to display the company's values and work dynamics without people around. In addition, some clients are now calling for on-site activities to close deals for several reasons, including security and communication measures. However, the company acknowledges the benefits of working from home to several professionals. This is why they need to define a hybrid configuration that works, and understanding the software professionals' needs and behaviors is the first step toward achieving this goal.

## 3.2 Data Collection

We applied three data collection techniques: questionnaires, documentary analysis, and observations. Note that our *dominant* data collection technique is a questionnaire. While most case studies are interview-focused, we used questionnaires to collect data from as many professionals as possible and obtain a representative snapshot of the case. The data collected from questionnaires were supported by the data emerging from documents and observations made through the company's official channels (e.g., Slack and email lists).

Regarding the questionnaire, we invited all software professionals in the company to answer the questions. However, for this case study, we are only interested in the answers from software professionals, i.e., those who work directly in software development activities, while the answers from other professionals (e.g., administrative roles) are stored for future work.

We used the company's channels to interact with employees and advertise the research, including the official email lists, Slack channels, and WhatsApp groups. The questionnaire remained open for two weeks (end of April 2023). During this period, we kept interacting with the software professionals, e.g., by posting the survey link and working to reach as many participants as possible.

While the questionnaire was being promoted, the professionals were constantly reminded that the survey was optional; however, it was important to increase the understanding of hybrid work in their company and the software industry in general. Data collection finished with 722 participants, from which 545 belong to our targeted population (e.g., software professionals) and 178 are employees in administrative roles (e.g., human resources and help desk).

In addition to the questionnaire, for documentary analysis, we collected available records that human resources and the help desk could share about the frequency to which participants were going to the office (e.g., clock-in reports and meeting room requests). As for observations, we focused on collecting data from relevant discussions about hybrid work triggered by professionals in the company's channels whenever the subject was mentioned. To document our observations, we created memos using a strategy based on netnography [15].

*3.2.1 Questionnaire.* To build the survey questionnaire, we first needed to explore how the company is handling post-pandemic hybrid work. Therefore, we interviewed three executives and two human resources employees and asked them about their current hybrid work structure. Based on the answers to this interview, we designed a questionnaire to anonymously collect software professionals' viewpoints on returning to the office on a regular basis.

We started the questionnaire with demographic questions about the participants, including their background, household details, transportation and commuting habits, role in the software team, years of experience, and time on the job, which are factors that were previously discussed to influence the perspective of software professionals working from home. Then, we asked participants about their work routine and their experience with the hybrid work model, how they feel when working remotely, and how they feel when working from the office, including benefits and limitations associated with both scenarios. We asked participants about their productivity and how the work configurations affect their work. We finished the questionnaire by asking participants about their frequency and motives to go to the office lately, their difficulties, and their preferences regarding hybrid work.

A pilot questionnaire was reviewed by a squad created in the company to discuss and understand hybrid work and validated with the participation of three senior employees, including software engineers and software designers. After consideration, some questions were modified or excluded from the questionnaire because our review and validation process revealed that they could confuse the employees or reduce the number of answers.

Thus, the questionnaire was modified to limit the closed-ended question about professional roles to include only the taxonomy used in the company, resulting in some software specializations, e.g., requirements analysts, being removed. Similarly, the options for diversity groups were adjusted to match those defined in the organization, including LGBTQIA+, Neurodivergence, and Person with Disability (PWD). Further, questions referring specifically to methods and tools used in the projects were excluded since some projects apply tools that, if mentioned, could risk the anonymity of participants and clients.

*3.2.2 Documents and Observations.* Data collected from documents obtained from human resources and IT support, as well as observation of discussions that happened in the company's channels, supported us in refining the data collected with the questionnaire and improved our understanding of software professionals' actions and behaviors. From human resources and IT, we obtained anonymous data about the frequency of people visiting the office and scheduling meeting rooms. Additionally, our observation was focused on discussions about people's interactions in the office while engaging in work or social activities.

### 3.3 Data Analysis

Both quantitative and qualitative data were obtained in this study. Therefore, due to the nature of these data, we applied two data analysis techniques. First, descriptive statistics [19] were applied to analyze the quantitative data collected from participants in the sample. Second, qualitative analysis was applied to open-ended questions in the questionnaire, document analysis, and notes from the observations based on three coding stages: line-by-line, focused, and theoretical coding [20].

*3.3.1 Quantitative Analysis.* We employed descriptive statistics [19] to quantitatively describe the key characteristics of our case. This involved dividing participants' responses into sub-groups using various statistical functions, such as means, proportions, totals, and ratios [17]. Descriptive statistics is a commonly used method for summarizing and presenting quantitative survey data [27]. In this process, we used Tableau to establish connections and comparisons between the frequency of office and remote work among software professionals and various sample characteristics, including gender, diversity, household configuration, software development experience, tenure, and team roles. We created interactive dashboards with visualizations, such as averages and distributions. The obtained aggregations allowed us to investigate specific groups of participants, such as those working in the office, sharing households with family, or having children. We adapted visualizations and explored different data combinations, and further, the qualitative data added depth to our case study analysis by providing insights into the attitudes and behaviors of software professionals.

*3.3.2 Qualitative Analysis.* Our qualitative analysis process involved three sequential phases [20]. Initially, we engaged in line-by-line coding and systematically explored open-ended responses to extract initial codes and develop data-driven concepts, as shown in Figure 1. Subsequently, we used focused coding and delved into the initial codes, identifying patterns and relationships and creating higher-level categories that encapsulated specific study aspects, as demonstrated in Figure 2. Finally, we conducted theoretical coding to iteratively refine these categories to gain a comprehensive understanding of the preferences of software professionals regarding hybrid work and how this work arrangement affects their activities.

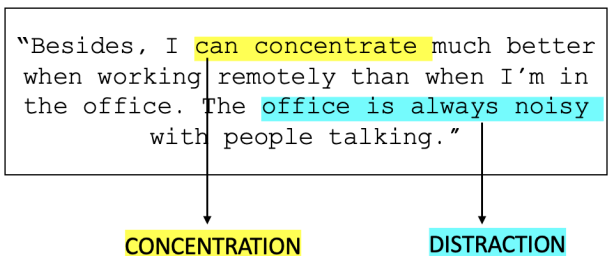

Figure 1: Example of line-by-line coding

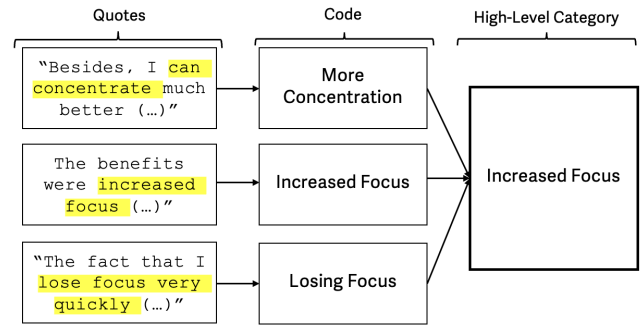

Figure 2: Building categories with focused coding

### 3.4 Ethics

No personal information about the participants was collected in this study (e.g., name, e-mail, or team) to maintain participants' anonymity. Questionnaires did not require any personal information. We only collected documents that were de-anonymized, and notes and memoing resulting from observing conversations in the company's channel did not include any identification of the employee, only sentences and phrases that were later merged into the pool of qualitative data.

## 4 FINDINGS

In this section, we present our main findings. We start by summarizing the characteristics of our sample of participants. Following this, we present the characteristics of software professionals who are working in the office regularly. Then, we discuss the reasons behind software professionals' attitudes towards opting for a hybrid work structure that fits their needs. Finally, we present the preference of professionals regarding hybrid work.

### 4.1 Demographics

We obtained 545 valid questionnaires with a highly diverse sample of participants, composed of software engineers from different profiles and backgrounds. They perform a variety of activities in software development and have experience in hybrid work in different ways. Our sample has an average of 10.6 years of experience working in software development (STD Dev= 7.5 years) and an average of 3.9 years working for the company that participated in the study (STD Dev= 4.3 years). Almost 75% of our sample is composed of professionals who live in households with no young children. Additionally, 46% of our sample is composed of programmers. Finally, our sample has a diversity of gender, ethnicity, and sexual orientation, which are important aspects discussed in recent studies on remote and hybrid work [11, 12, 33]. Table 1 presents a full overview of our participants, including living arrangements, main commuting type, hybrid configuration preferences, and perceived productivity.

**Table 1: Demographics**

| | Participants Profile | |
|---|---|---|
| Gender | Male | 398[a] individuals |
| | Female | 131[b] individuals |
| | Non-binary | 3 individuals |
| | Prefer not to answer | 13 individuals |
| Ethnicity | White | 352 individuals |
| | Mixed-Race | 137 individuals |
| | Black | 34 individuals |
| | Asian | 4 individuals |
| | Prefer not to answer | 18 individuals |
| Underrepresented Group | Neurodivergent | 87 individuals |
| | LGBTQIA+ | 63 individuals |
| | Person with Disabilities | 12 individuals |
| Living Arrangement | Family | 475 individuals |
| | Alone | 68 individuals |
| | Friends | 8 individuals |
| | Other | 4 individuals |
| Parents (Have Kids) | No | 407 individuals |
| | Yes | 138 individuals |
| Main Commuting Type | Private Vehicle | 209 individuals |
| | Apps | 120 individuals |
| | N/A | 80 individuals |
| | Public Transportation | 71 individuals |
| | Airplane | 38 individuals |
| | Bike | 14 individuals |
| | Ride | 10 individuals |
| | Walk | 3 individuals |
| Role | Developer | 250 individuals |
| | QA | 138 individuals |
| | Designer | 67 individuals |
| | Manager | 47 individuals |
| | Data Scientist | 27 individuals |
| | Other | 16 individuals |
| Experience | 0-2 Years | 59 individuals |
| | 3-5 Years | 105 individuals |
| | 6-10 Years | 155 individuals |
| | 11-15 Years | 101 individuals |
| | 15-20 Years | 68 individuals |
| | 20+ Years | 57 individuals |
| Time in Job | 0-2 Years | 315 individuals |
| | 3-5 Years | 122 individuals |
| | 6-10 Years | 48 individuals |
| | 11-15 Years | 39 individuals |
| | 15+ Years | 21 individuals |
| Current Hybrid Configuration | Office mode (4-5 times a week) | 9 individuals |
| | Office-first (2-3 times a week) | 53 individuals |
| | Office-remote Mix (1 a week) | 59 individuals |
| | Remote-first (eventually) | 241 individuals |
| | Remote-mode (Not Going) | 183 individuals |
| Preferred Hybrid Configuration | Office mode (4-5 times a week) | 3 individuals |
| | Office-first (2-3 times a week) | 41 individuals |
| | Office-remote Mix (1 a week) | 22 individuals |
| | Remote-first (eventually) | 336 individuals |
| | Remote-mode (Not Going) | 143 individuals |
| Perceived Productivity | Way More Productive | 185 individuals |
| | More Productive | 176 individuals |
| | Same Productivity | 156 individuals |
| | Less Productive | 28 individuals |
| | Way Less Productive | 0 individuals |
| Weekly Overtime | No overtime | 269 individuals |
| | 0-2 Hours | 184 individuals |
| | 2-5 Hours | 75 individuals |
| | 5+ Hours | 17 individuals |

Notes: [a] 2 transgender men. [b] 2 transgender women

## 4.2 Who are the Software Professionals working in the Office?

Regarding how software professionals are experiencing hybrid work, our findings demonstrated that most software professionals are opting to work in schemes that are close to remote (e.g., remote mode and remote-first) and going to the office on rare occasions, while only a small number of individuals developed a routine that includes working in the office regularly, e.g., weekly. In our case, we identified that:

- 44% of the participants (241/545 individuals) are working on remote-first mode and going to the office eventually at least once a month.
- 33% the participants (183/544 individuals) are working on fully remote work and have never returned to the office after the pandemic restrictions were lifted.
- 11% (59/545 individuals) the participants are working on an office-remote mix mode and going to the office regularly, usually once a week.
- 10% (53/545 individuals) the participants are working in an office-first mode, as they are working in person more regularly, something like two or three times a week.
- 2% (9/545 individuals) the participants are back to how they were used to working before the pandemic and going to the office mostly every day in an office mode configuration.

This classification of hybrid work arrangements is consistent with our research findings, as our analysis is based on the individual perspectives of software professionals regarding hybrid work. In this context, even when the majority of team members opt for office-based work, they need to maintain a hybrid structure to accommodate those who prefer arrangements that are more remote in nature. Conversely, the reverse scenario is equally valid. If an individual chooses to work in an office-mode arrangement, other team members may opt for different setups; thus, from a team perspective, they collectively operate within a hybrid work context.

We thoroughly examined the unique attributes of each hybrid arrangement and observed that certain participant attributes, such as gender, ethnicity, experience level, and job tenure, do not exert a significant influence on their choices regarding hybrid work configurations. In contrast, our findings suggest that factors such as professional role within the team, living arrangements, and commuting methods have a subtle yet discernible influence on professionals' decisions regarding hybrid work. In section 4.3, we further explore the reasons that might be influencing these factors.

Regarding the roles and responsibilities within the software project, our results indicate that software testing professionals and software project managers are more prone to adopting an office-centric routine with weekly in-office commitments. In contrast, software designers and data scientists predominantly opt for remote work, making occasional office visits. The distribution of programmers between the office and remote work groups is relatively balanced. Figure 3 illustrates these findings.

Upon examining the commuting patterns, we noticed that professionals residing at considerable distances from the office who rely on air travel as their primary commuting method form a minority among those who frequently visit the office. A similar trend emerges among professionals who use eco-friendly commuting options such as biking or walking. On the other hand, we noted a higher frequency of office attendance among those who rely on various car-related options, including personal vehicles, ride-sharing apps, or public transportation. Figure 4 demonstrates these findings, and a broader range of insights on commuting is presented in section 4.3.

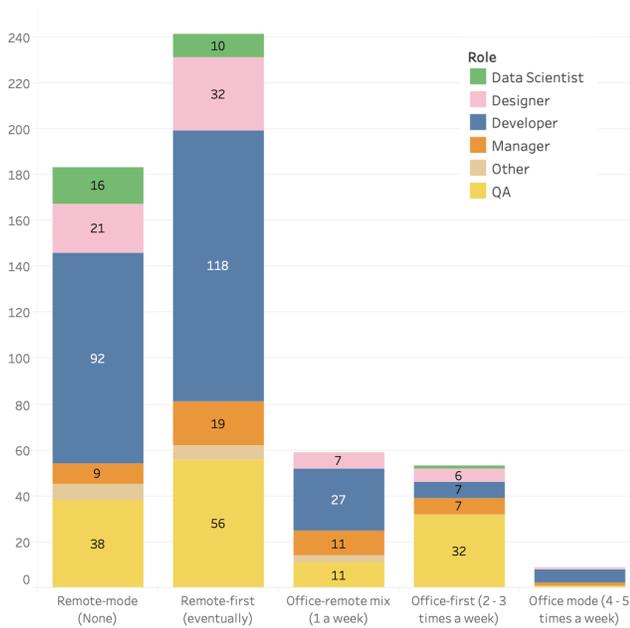

Figure 3: Hybrid Configurations Based on Roles

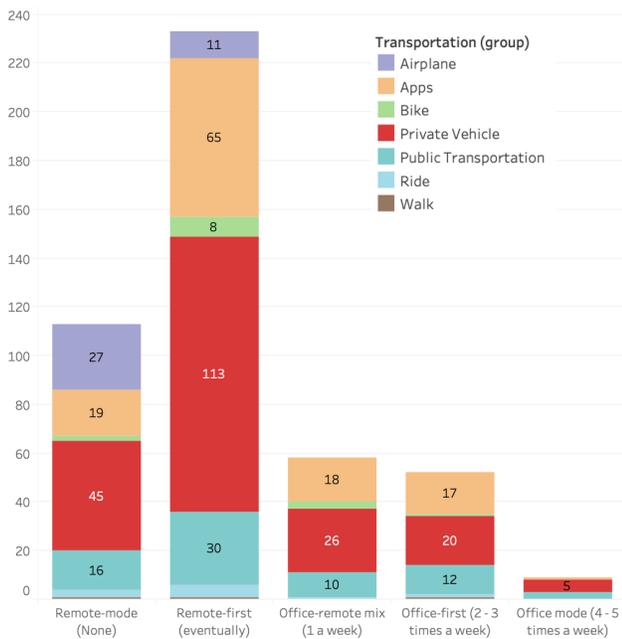

Figure 4: Hybrid Configurations Based on Commuting

In the context of household configurations, our analysis reveals that software professionals who reside in households without the presence of family members tend to visit the office less frequently compared to those who share their living space with family. This remote-focused group encompasses individuals living alone, with friends, or in other arrangements not involving family members.

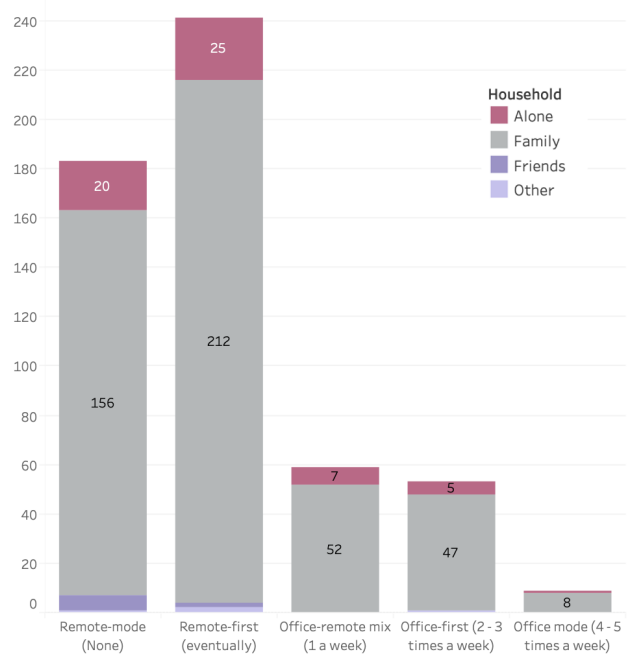

Figure 5: Hybrid Configurations Based on Household

Figure 5 shows these findings. Initially, this discovery may suggest that some professionals encounter challenges when blending their work responsibilities with family life. However, a closer look at the qualitative data collected in our study reveals that concentration and focus appear to be the primary factors at play in this context, as elaborated upon in section 4.3.

### 4.3 What factors influence the decision-making process of software professionals when it comes to selecting hybrid work arrangements?

We have pinpointed six primary factors that drive software professionals in hybrid work settings to visit the office. These factors may be rooted in personal preferences or of a professional nature, arising from project or team requirements. In general, nowadays, software professionals commonly find themselves in the office for the following reasons (illustrated in Table 2):

- *Social Interaction*: A substantial portion of professionals expressed a desire for more frequent social interactions with their colleagues and team members. They perceive the office as a space where these interactions can be amplified, whether through informal gatherings or social activities organized by the company.
- *Ergonomics Improvement*: While many software professionals have adapted their home offices to suit their work requirements, some still look to the office to enhance their daily ergonomics, aiming for better work equipment such as desks, chairs, and air conditioning, among other amenities.

- *Team Meetings*: To a certain extent, software professionals choose to frequent the office because the team, as a collective unit, has decided that certain meetings and activities are more effectively conducted in person to leverage direct communication and smoother coordination.
- *Infrastructure Aspects*: Software professionals may return to the office due to challenges impacting their work related to the need for support and essential infrastructure. This can involve specific technology to complete tasks that are only available in the office or seeking specialized support for equipment problems.
- *Management Requests*: For some professionals, their regular presence in the office is not a matter of personal preference but rather a result of explicit requests from their managers or technical leaders, who require their in-person presence for specific activities.
- *Client Demands*: While the organization and the team may establish guidelines for hybrid work configurations, there are instances where certain professionals are required to work from the office to meet specific client demands, e.g., the needs from product owners, business partners, project sponsors, and other stakeholders.

Upon examination, we can categorize the factors driving software professionals back to the office into three distinct groups: occasional self-driven incentives, project and team-related requirements, and institutional regulations. Factors such as the need for social interaction or the pursuit of improved ergonomic conditions are primarily fueled by personal preferences and individual motivations. On the other hand, team meetings and infrastructure considerations are aligned with collective goals, but they are not mandatory. In contrast, management requests and client demands represent rules driven by the institutional hierarchy, imposing a compulsory return to the office.

Additionally, as per our analysis, we conceptualize the hybrid work environment as a continuum spanning from working in office-mode work to remote-mode. Therefore, to gain a comprehensive understanding of the factors influencing software professionals in their choice of hybrid work arrangements, we must also delve into what motivates them to work remotely. In this context, we have identified five key factors (illustrated in Table 2):

- *Commuting Avoidance*: The daily commute, once a standard aspect of work life, has become a source of concern for software professionals due to extended and exhausting commutes, traffic problems, parking difficulties, and concerns related to public transportation. Currently, there is a common belief that the time spent on commuting could be better allocated to project-related tasks.
- *Family Time*: What was previously perceived as a potential distraction, having family members around more often, has now become a significant aspiration among software professionals. This heightened appreciation for family time has driven many to explore flexible work arrangements, recognizing that spending quality time with family offers emotional support, stress reduction, and contributes to overall well-being.
- *Increased Focus*: Having adapted to maintaining concentration and focus in their home offices while sharing the space with family, software professionals now find traditional office environments to be more interruptive and distracting, largely due to noisy conversations. As a result, they often seek quieter, more focused spaces at home for their work activities.
- *Work-life Balance*: The pandemic underscored the importance of work-life balance as individuals have to juggle their professional responsibilities with personal well-being. For software professionals, the flexibility inherent in hybrid work arrangements provides an opportunity to strike a more harmonious balance between their work commitments and other essential aspects of life, such as education, healthcare, and leisure.
- *Comfort and Privacy*: Software professionals value the comfort and privacy provided by remote work settings, allowing them to personalize their workspace to suit their preferences, contributing to a more enjoyable and motivating work environment while also enhancing their ability to handle personal matters and discuss sensitive project information discreetly.

Some of the benefits driving software professionals to work in remote-like modes are linked to the broader characteristics discussed in Section 4.2. Commuting is no longer an option, particularly as many professionals now reside farther from the office than before the pandemic. Additionally, the balance between focus and distractions sheds light on why individuals living alone often opt for remote work, given the high level of distractions encountered in office environments and their preference for the comfort and privacy of home offices. Overall, the factors influencing the decision-making process of software professionals underscore the prevailing preference for hybrid work setups that prioritize a more remote-oriented approach, which is perceived as conducive to increased productivity.

### 4.4 What is the relation between hybrid work and the *perceived* productivity of software professionals?

As shown in Table 1, the software professionals in our case study mostly reported that hybrid work has had a favorable effect on their productivity. Their overall perception is that they have been able to convert time that was previously spent on commuting, long meetings, informal interactions, and extended breaks into dedicated work hours. Furthermore, the enhanced focus derived from remote-oriented environments has contributed to more efficient task completion, resulting in reduced delays and an overall boost in productivity.

An important factor to consider in evaluating the productivity of software professionals in hybrid work settings is that, despite many reporting increased productivity, a substantial group also acknowledges working overtime, with some even exceeding five extra hours per week. Our analysis of the qualitative data obtained from participants revealed that this overtime often occurs organically, driven by the blurred boundary between home and work life.

Table 2: Quotations Supporting Findings

| Factor | Quotes |
| --- | --- |
| Social Interaction | "the most rewarding experience was reuniting with my colleagues, sharing meals, enjoying moments together, and having coffee breaks." (P246)<br>"I really enjoy interacting with people, and this is not possible in remote work (without establishing direct connections)." (P251)<br>"The day takes on a more pleasant tone as we reconnect with friends, engage in conversations, enjoy coffee breaks, and share meals" (P437) |
| Ergonomics Improvement | "[office] air conditioning providing an ideal working environment." (P388)<br>"The only issue is that my chair [at home] is quite uncomfortable." (P175)<br>"The overall experience is positive, yet challenges do appear, like bad weather (excessive heat) and less-than-ideal workspace conditions (plastic table and limited space)." (P408) |
| Team Meetings | "Working alongside your team members facilitates communication." (P461)<br>"Team interaction is enhanced [in the office], making project equipment organization more efficient." (P481)<br>"Holding in-person meetings occasionally can be more productive.." (P011) |
| Infrastructure Aspects | "Each team member possessed specific equipment, and we had to run a test with 5 devices connected to a single machine to address a critical bug." (P535)<br>"There were factors that hindered productivity, such as power outages and internet disruptions." (P489)<br>"Difficulty can arise when there's an issue that requires taking the computer to the office for repair." (P484) |
| Management Requirements | "Working in the office is the choice of micromanaging managers, which can be frustrating due to someone constantly overseeing your work" (P175)<br>"there is room for improvement in leadership to foster a more comfortable and empowering environment for the employees." (P212)<br>"If I move up to a more leadership role, I would recognize the importance of being more present in the office." (P300) |
| Client Demands | "I've never worked in person, except for two days, one of which was solely for meetings with the client." (P424)<br>"when the client is on-site or when the team requires strategic meetings." (P476)<br>"The week I worked in the office was due to a client visit, which led to modifications in the project's routine." (P005) |
| Commuting Avoidance | "there are no issues at home that are WORSE than having to endure 3 hours of daily traffic to cross the city from one end to the other." (P107)<br>"The fear of being robbed or mugged is a genuine worry. I've personally experienced a mugging on my way home from work, and both my sister and wife have had similar encounters during their commutes." (P371)<br>"And I felt truly exhausted due to the commute." (P049) |
| Family Time | "The concern for some of my family members would prevent me from focusing 100% on work [in the office]" (P410)<br>"I would be demotivated because I had to disrupt the established routine for my autistic children." (P130)<br>"I had a child during the pandemic, and now I can be closer to them." (P153) |
| Increased Focus | "I struggle to focus in environments where people are chatting with each other, coming in and out of the premises." (P165)<br>"I always lost focus because there were too many distractions in the office." (P176)<br>"I can't concentrate for even 30 minutes when I'm in the office because the noise is incessant." (P175) |
| Work-Life Balance | "I can balance my work activities well with my personal and family life.." (P323)<br>"The main benefit is improved quality of life (...) I can sleep better, go to the gym, study for my master's degree, etc." (P419)<br>"The clear benefits include the flexibility to align work hours with personal needs, such as medical appointments going to the gym." (P050) |
| Privacy and Comfort | "Air conditioning [in the office]. I suffer from allergies due to air conditioning." (P192)<br>"greater ease and privacy for discussions on sensitive topics in 1:1s and critical meetings" (P478)<br>"I have a strong aversion to the cubicle-style work setup that continues to endure. I believe this design harks back to a time decades ago." (P529) |

Figure 6 illustrates the connection between perceived productivity and reported overtime work.

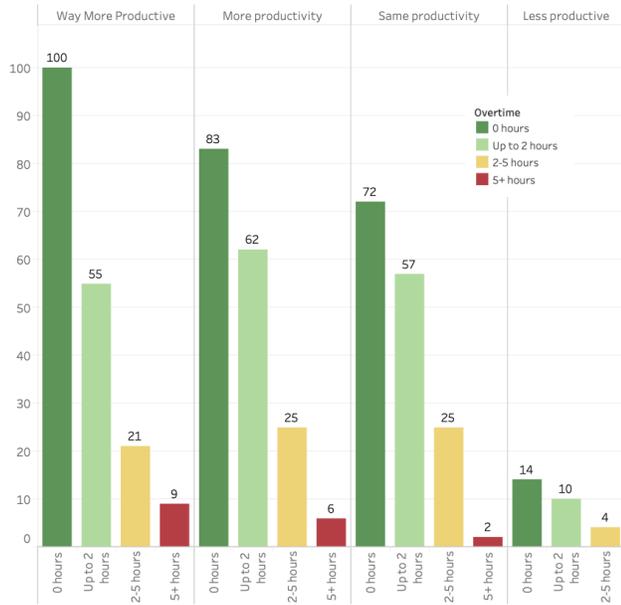

**Figure 6: Perceived Productivity and Overtime**

## 4.5 What are the preferred hybrid work configurations among software professionals?

As highlighted in our findings detailed in Section 4.3, the decision to work in a more remote or office-centric mode among software professionals is not always an individual choice. This decision can be shaped by collective agreements within the team or institutional and hierarchical rules that mandate the preferred hybrid configuration. For example, as discussed in Section 4.2, software testing professionals are frequently found working from the office. However, we identified that management requests typically drive them to do so. This tendency is likely related to the significant impact that software quality has on project costs and timely deliveries.

In this context, we explored the preferences of software professionals regarding hybrid work and compared them with their current work configurations. The findings, as illustrated in Figure 7, indicate that the majority of software professionals express a preference for remote and hybrid-first work modes, even among those currently engaged in remote-mix or office-first arrangements. However, there is also a segment of professionals presently in more remote-oriented setups who would actually prefer to transition to an office-based configuration. These observations reinforce the complexity of the current hybrid work landscape in the software industry, where determining the ideal hybrid configuration is a challenging task influenced by various individual, team, and organizational factors.

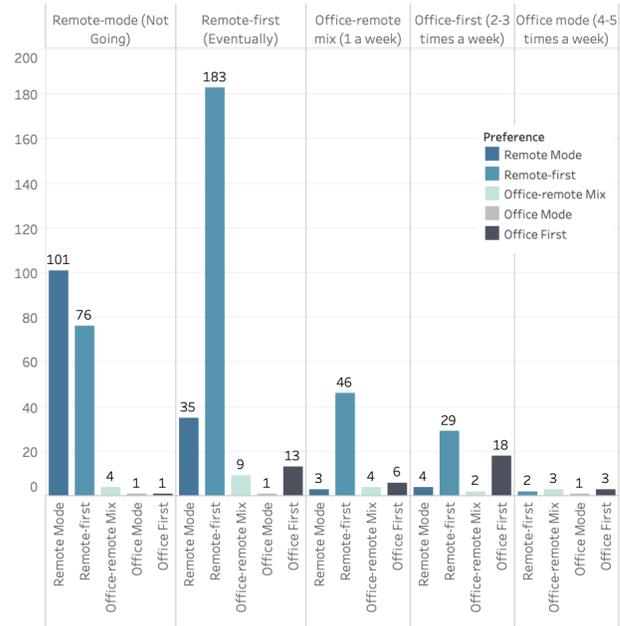

**Figure 7: Perceived Productivity and Overtime**

## 5 DISCUSSION

Hybrid work has firmly established itself as a prevailing reality within the software industry [38]. This scenario was accelerated in the post-pandemic era in response to the increasing demand for flexibility among software professionals to maintain the benefits experienced when fully remote work was the norm [1, 11, 31]. Consequently, we observe today that the line between traditional office-based work and remote work has become increasingly blurred [39].

Working in hybrid environments means embracing a range of work configurations that incorporate varying degrees of in-person and remote activities [37]. In this evolving landscape, numerous software companies have embraced a voluntary hybrid work model, allowing their employees to select the work environment that aligns with their individual preferences [40]. However, our research underscores that this decision-making process is complex, extending beyond individual needs and encompassing considerations at the team, project, client, and organizational levels.

Our research highlights the importance of achieving a harmonious balance to unleash the potential of hybrid work fully. This equilibrium is vital for meeting the flexibility expectations of software professionals while providing the necessary structure to drive team productivity and collaboration, all in alignment with client and business needs. In this context, enhancing the attractiveness of office spaces reduces the need for compulsory attendance measures that may lead to dissatisfaction among software professionals. Simultaneously, providing essential infrastructure to support remote activities remains a fundamental consideration. Most significantly, it is crucial to acknowledge that hybrid work is a collective construct involving multiple software professionals operating in diverse hybrid configurations, and ensuring these configurations harmonize effectively is paramount to their success.

### 5.1 Implications for Practitioners

Software companies aiming to increase office-mode work should prioritize the benefits of social interactions, robust infrastructure, and engaging activities. These actions can mitigate the downsides of in-office work and attract more software professionals to the physical workplace. Fostering spontaneous interactions through well-planned common areas and events contributes to a vibrant office culture. Simultaneously, managing distractions and creating an office layout that promotes focus is crucial. Additionally, nurturing a sense of belonging and community through a robust organizational culture can significantly increase the desirability to be in the office.

Software project managers should avoid depending solely on mandatory rules to encourage office participation when required. Embracing the advantages of remote work is essential, and they should acknowledge the positive impact of remote-mode configurations on their teams while also motivating professionals to actively engage in key in-person activities that genuinely require physical presence. Additionally, they have a crucial role in ensuring the success of the hybrid structure, which demands that everyone in the team feels included and collaborates effectively. By promoting a culture of flexibility and understanding individual needs, project managers can strike the necessary balance to effectively navigate hybrid work.

Software professionals should maintain open and transparent communication with their companies and teams about the impact of hybrid work on their professional and personal lives. They should highlight the positive aspects of hybrid work, such as increased focus, reduced commuting time, and improved work-life balance. However, they should also be willing to attend in-person key activities when necessary to support the team. Additionally, it is crucial to establish clear boundaries between work and personal life to prevent excessive overtime and maintain a healthy balance. Open dialogue and well-defined expectations are essential for maximizing the benefits of hybrid work while avoiding potential pitfalls.

### 5.2 Implications for Researchers

Our research has revealed various practical implications related to post-pandemic hybrid work and its consequences for software teams. However, it is evident that there is much more to explore within this context. Currently, our focus is on further examining the extensive data gathered from our case study. We aim to gain a more comprehensive understanding of how hybrid work influences specific groups, such as software testers, who seem to be engaging in more in office-mode activities, and software professionals from underrepresented groups, as many factors driving remote work have particular relevance to them. Moreover, we are exploring the broader ramifications of hybrid work on the overall well-being and health of software professionals, as many identified factors have a direct impact on their quality of life.

Researchers can leverage our study for further investigations into hybrid work configurations within software engineering. Currently, there is a pressing need to determine the most suitable hybrid work configurations for various project types and to understand how these configurations impact agile methodologies. Additionally, identifying the tools and technologies that can effectively support hybrid work is essential, and theories and strategies must be developed to assist software companies in effectively planning and implementing this work model. By expanding on these areas, researchers can contribute to enhancing the efficiency and adaptability of hybrid work models in the ever-evolving landscape of software development.

### 5.3 Threats to Validity

Case studies, while valuable in exploring real-world phenomena, carry inherent threats to validity. One key concern is related to construct validity, as the interpretation of data and the definitions of concepts may influence the results. In this sense, we acknowledge that our findings might have been influenced by our interpretation of the field. To address this potential source of bias, we employed a triangulation approach, leveraging data from multiple sources (questionnaires, documents, and observations) to enhance the reliability and validity of our findings. Furthermore, we strongly relied on direct quotes from the participants and consistently compared and contrasted codes and emerging themes to describe our findings.

Moreover, in terms of external validity, it is important to emphasize that findings derived from case studies are not intended for direct generalization to broader populations. Instead, case studies provide context-specific insights. Hence, while our specific findings may not be statistically generalizable, our discussions offer valuable insights that can be adapted and applied in various settings. To facilitate knowledge transfer, we have provided as much contextual information about our case and participants as our ethical agreement with the company permits. We expect that this enable other researchers and practitioners to derive insights from our work and apply them within their contexts.

## 6 CONCLUSION

During the post-pandemic era, the global workforce experienced a profound transformation, and the idea of hybrid work became a central topic of discussion within software companies. Our objective in this study was to investigate the reception of hybrid work arrangements within the software industry by exploring the experiences of software professionals and the various factors influencing their preferences. We aimed to shed light on the complex dynamic between individual preferences, team dynamics, and organizational strategies in the context of post-pandemic hybrid work.

Our industrial case study provided a rich understanding of the multifaceted landscape of post-pandemic hybrid work based on the experiences of a diverse group of 544 software professionals. Our findings have highlighted the complex nature of hybrid work adoption, underscoring that it is not a simple individual decision but rather a convergence of professional, team, project, client, and organizational factors. Our empirical evidence has underscored that, in the present context, there is no one-size-fits-all approach to hybrid work in the software industry.

The implications of our research have wide-reaching significance in software engineering. As the adoption of hybrid work models continues to grow, it is essential to acknowledge the balance among flexibility, collaboration, work-life balance, and productivity. In this mosaic, software professionals should advocate for their preferences and communicate how hybrid work arrangements can enhance their overall well-being, productivity, and job satisfaction. Meanwhile, software companies should strive to cultivate an organizational culture that effectively encourages professionals to engage in meaningful social activities while simultaneously mitigating office distractions. Lastly, software engineering researchers should persist in exploring the nuances of hybrid work to provide guidance to practitioners as they navigate the multifaceted challenges associated with this evolving work context.